# Applicability of DUKPT Key Management Scheme to Cloud Wallet and other Mobile Payments

Amal Saha
Email: amal.k.saha@gmail.com
Tata Institute of Fundamental Research (TIFR), INDIA

Sugata Sanyal
Email: sugata.sanyal@tcs.com
Research Advisor
Tata Consultancy Services, Mumbai, INDIA

**ABSTRACT**
After discussing the concept of DUKPT based symmetric encryption key management (e.g., for 3DES) and definition of cloud or remote wallet, the paper analyses applicability of DUKPT to different use cases like mobile banking, NFC payment using EMV contactless card and mobile based EMV card emulation, web browser based transaction and cloud or remote wallet.

Cloud wallet is an emerging payment method and is gaining momentum very fast. Anticipating that the wallet product managers and security specialists may face these questions from different stakeholders, the authors have addressed applicability of DUKPT to cloud wallet use case quite elaborately. As per knowledge of the authors, this topic has been analysed and discussed for the first time.

## 1. INTRODUCTION
Encryption is very widely used payment systems to ensure confidentiality of payment and associated data in transactions. Since encryption algorithms (3DES or TDES, AES, etc) are standardised and published, it is important to manage the key very securely. Distribution of key is an important step while setting up encryption ecosystem and this has impact on the way business process is executed during actual payment transaction, and management and efficiency of operations.

Derived Unique Key Per Transaction (DUKPT) is an approach for managing encryption keys of symmetric-key algorithms like 3DES, AES, etc in a card payment environment. It is not an encryption algorithm. In this scheme, encryption uses a derived key that once used in a transaction is not used again for a second transaction.

Predecessor of DUKPT based key management was an approach where a unique fixed key is shared between a payment terminal and the backend service (e.g., of acquirer). This is called fixed key scheme. Another predecessor, called Master/Session key management scheme, uses a pre-shared Key Encrypting Key (called "Master") to encrypt a data key (called "Session" key) generated randomly and communicated insecurely. Data key is used for encrypting data to be exchanged between two entities. These techniques were created in the days before asymmetric encryption techniques. With millions of devices in circulation, this can be a daunting task to manage the operation of injection of unique fixed key or key encrypting key per device and ongoing secure storage of the large number of such keys on the server side of the processor. While these are simple, communicating the pre-shared fixed key or Key Encrypting Key and maintaining them are challenging.

Traditionally encryption key has been stored in a secure cryptographic device with various levels of certification (e.g., Hardware Security Module, EMVCo or PCI certified POS terminal) [2, 3,11, 12 ].

DUKPT is widely used to encrypt electronic commerce transactions on point-of-sales (POS) terminals for card payments and also ATM machines for banking.

VISA [8] was the inventor of DUKPT and it has been used with POS terminals in North America, but adoption has been significant in other regions and for other applications. In short, DUKPT has become very important in the payments space. Various related security aspects have been discussed in [13, 14, 15].

This paper is about applicability of DUKPT key management scheme to a type of mobile payment called cloud or remote wallet which is defined subsequently, and also other use cases.

## 2. DUKPT METHODOLOGY AND DEFINITION OF CLOUD WALLET

### 2.1 WORKING OF DUKPT

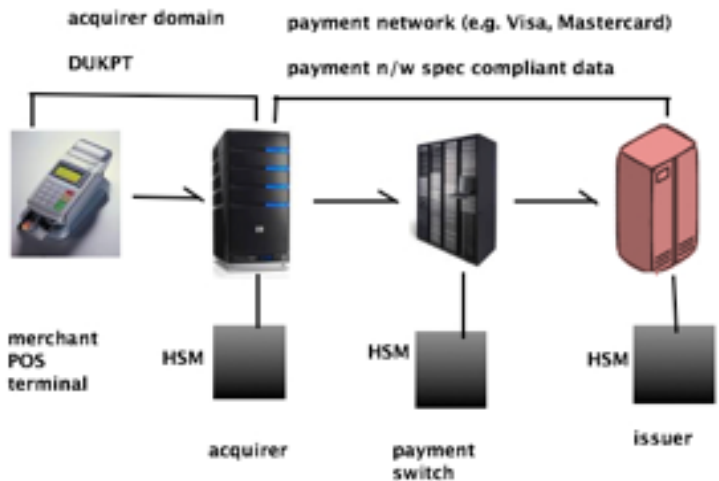

Figure 1 : DUKPT in payment

DUKPT key management is typically used for 3DES (or AES) keys used in POS terminals (Figure 1). This key management scheme may be used separately for encrypting card PIN (personal identification number) block and also for encrypting payment transaction data. These techniques have been employed in designs of many POS terminals including mobile POS terminals [4]. The point-to-point encryption is used between the POS terminal and a subsystem in acquirer processor. The data communication between the acquirer processor and the payment network (e.g., Visa, MasterCard) may not use DUKPT and may instead use fixed-key or Master/Session key management scheme.

There could be variation of the flow depicted in Figure 1. For example, there could be a gateway (not shown in Figure 1) between the POS terminal and the acquirer processor, and only the POS terminal and the gateway use DUKPT and other systems use other key management schemes. Other variations are also possible.

There are three levels of keys in DUKPT - BDK, IPEK and Transaction Keys. A brief summary of initial setup and transaction-time processing is given below.

**Base Derivation Key (BDK)** - a 3DES master key owned by the acquirer. The same BDK can be used for a large number of terminals, perhaps all the terminals that the provider supplies or for serial number range or a model of terminals. If the BDK is owned by entity other than the acquirer, it will need to be distributed to the acquirer. It needs to be distributed to any other entity involved in generating IKEYs or IPEKs.

**Initial Key (IK, IKEY or IPEK (initial PIN encryption key))** - a 3DES key that is unique to a terminal, and is calculated from BDK and the initial Key Serial Number (KSN) which is unique to the terminal. This key is used to initiate the sequence of transaction keys. It is discarded by the terminal after use. The IKEY/IPEK is installed into the terminal. The IKEY is recreated on the fly by the acquirer at the time of processing transactions from the terminal to derive the same transaction key that the terminal used to encrypt the data.

**Transaction Keys**- generated within the terminal. When the IPEK is installed in the terminal, it calculates up to 21 "Future Keys" or Transaction Keys. These keys will be used in the encryption of future transactions. When the initial batch of Future Keys has been derived, the IKEY is no longer required, and is deleted by the terminal. Keys for PIN encryption, data encryption, etc are derived from the transaction key or future key. Each transaction data is protected thru use of unique transaction key. When the encrypted data is received by the acquirer, the acquirer will derive the same transaction key or future key using exactly same process that the terminal used to derive the encryption key. The KSN is modified by incrementing the Transaction Counter. The DUKPT terminal sends its encrypted data and the KSN, together with other transaction data, to the acquirer.As a Transaction Key is used, it is deleted by the terminal and replaced by a new key.

## 2.2 DEFINITION OF CLOUD/REMOTE WALLET

Mobile wallets are becoming popular and both proximity (NFC) [9] and cloud based wallet [9] payments are gaining momentum. For proximity payment, EMV contactless payment specifications [6] would apply and DUKPT key management scheme for encryption of transaction data between the NFC-enabled POS terminal and the acquirer processor system is already supported.

Remote or cloud wallet refers to the scenario where payment details (payment instruments like card, bank accounts, etc), shipping address, etc are stored on the cloud (i.e., server) and for online purchase when the checkout process is initiated by the user on the mobile device or other end-point device (e.g., laptop), the server actually initiates the actual payment transaction with respective intermediary (e.g., bank's system, payment gateway, etc).

## 2.3 REVIEW OF CURRENT DUKPT USE CASES IN PAYMENTS AND OBJECTIVE OF THE PAPER

DUKPT is being used in standard contact based POS terminals (e.g., the ones used for in-store payment) [5] and other POS terminals that support contactless EMV card [6] and mobile based on EMV card emulation [7]. DUKPT is also supported in mobile POS terminals (e.g., the ones used by Pizza delivery boy to accept card payment) [4].

Gemalto [1], a leading mobile banking product vendor, reported DUKPT key management for 3DES encryption operation on mobile SIM card, for mobile banking. In this case the SIM card also known as Secure Element is a FIPS 140-2 [2, 3] certified hardware, with certification at various levels, Level 2 being minimum.

POS terminals attached to mobile phone of the merchant (e.g., the ones used by Pizza delivery boy to accept card payment) also started using DUKPT key management scheme [4]. NFC based wallets already use DUKPT for communication between the POS terminal and the acquirer processor.

Often the product managers of cloud wallet would be asking themselves and their information security specialists if they could enable DUKPT key management scheme for cloud or remote wallet as well.

This paper addresses this question, i.e., applicability of DUKPT key management scheme to the case of cloud wallet payment and also other mobile payment use cases.

## 3. APPLICABILITY OF DUKPT TO CLOUD WALLET PAYMENT and OTHER MOBILE PAYMENTS — CHALLENGES

It may be noted that DUKPT requires cryptographic relationship between the client-side device (e.g., browser on smartphone or laptop) and the backend service. One consequence is that DUKPT cannot be used in a secure way for transaction between web browser and the backend service because one cannot store IPEK securely on browser to establish that cryptographic relationship.

It has already been highlighted that NFC payments with mobile phones comply with EMVCo contactless [6] specification and DUKPT is automatically supported by the compliant contactless POS terminals.

DUKPT requires existence of a secure device for secure storage of the terminal's unique initial Key Serial Number (KSN). The initial KSN which is unique to the terminal is typically injected by the device manufacturer in its facility. Cloning of IPEK would lead to cloning of terminal for a given BDK (base derivation key).

Often IPEK is generated by the payment processor at its facility with the help of terminal's unique initial KSN provided by the device manufacturer and BDK in processor's hardware security module (HSM). Cloning of IPEK injected in a terminal will lead to cloning of the terminal and this must be prevented. This requires the terminal to be a secure cryptographic device and FIPS 140-2 certification at various levels or other certification by PCI or EMVCo is critical.

The cloud wallet product managers who would like to build DUKPT based encryption to enhance sales would perhaps think of encrypting every transaction that originates at the mobile device with a unique transaction key and which is processed by the backend service in the cloud wallet system.

In a mobile device, Secure Element (SE) [10] is the component that could be used for IPEK injection. It may be noted that SE has different form factors, namely, a) UICC or SIM card, b) hardware embedded SE or c) detachable micro-SD card. For a cloud based wallet which would like to use DUKPT for encrypting a checkout request on a mobile device using 3DES encryption algorithm, the only place to do so would be an SE. Most common SE form factor is SIM card. Challenge with the SIM card as cryptographic processor supporting storage of the IPEK is that the solution is highly dependent on the mobile network operator. The solution and challenges here are similar to the scenario of mobile banking that uses DUKPT [1]. Challenge with micro-SD card acting as cryptographic processor is that distribution of the SE has to be done by the wallet solution provider and the user has to carry that for any transaction and the user experience goes for a toss. As for embedded SE, there is a unique challenge — the mobile device manufacturer must collaborate with the wallet service provider to give access to the embedded SE.

Assuming that a solution that creates dependency on the mobile network operator or the wallet service provider or the device manufacturer, is not desirable to the mobile user, use of DUKPT for cloud wallet payment can be ruled out.

## 4. CONCLUSIONS

The concept of DUKPT based symmetric encryption key management (e.g., for 3DES) and cloud or remote wallet have been defined. Applicability of DUKPT to different use cases like mobile banking, NFC payment using EMV contactless card and mobile based EMV card emulation, web browser based transaction and cloud wallet have been discussed.

Cloud wallet is a hot topic in payment space and gaining momentum very rapidly. Since the wallet product managers and security specialists may someday face these questions, the authors have addressed applicability of DUKPT to cloud wallet use case quite elaborately. As far as knowledge goes, this topic has been analysed and discussed for the first time by the authors.

In future, adoption of DUKPT based key management for appropriate use cases illustrated in the paper, would go up in emerging payment solutions including mobile payments. Furthermore, mobile wallet product and solution managers would review appropriateness of DUKPT because not all use cases can leverage DUKPT in view of cryptographic requirements imposed on the overall ecosystem by DUKPT key management scheme.